\documentclass[twocolumn,twoside,slac]{revtex4}
\usepackage{graphicx}
\usepackage{fancyhdr}
\def\mrad{\mbox{mrad}}

\def\veps{\varepsilon}
\def\vepsb{{\bf \varepsilon}}
\def\bfr{{\bf r}}
\def\bfdr{{\Delta\bfr}}
\def\bfq{{\bf q}}
\def\bfdq{{\Delta\bfq}}
\def\bfR{{\bf R}}

\def\bfdR{{\Delta\bfR}}
\def\bfs{{\bf s}}
\def\hbfs{{\hat\bfs}}
\def\hbft{{\hat{\bf t}}}
\def\bfw{{\bf w}}
\def\hbfw{{\hat\bfw}}
\def\bJ{{\bf J}}
\def\bV{{\bf V}}
\def\bfp{{\bf p}}
\def\Da{{\Delta\alpha}}
\def\Db{{\Delta\beta}}
\def\Dg{{\Delta\gamma}}
\def\angu{{\psi}}
\def\angv{{\vartheta}}
\def\tanu{{\tan{\angu}}}
\def\tanv{{\tan{\angv}}}
\def\um{{\mu\mbox{m}}}

\begin{document}

\title{Sensor Alignment by Tracks}

%

\author{V. Karim\"aki, A. Heikkinen, T. Lamp\'en, T. Lind\'en}
\affiliation{Helsinki Institute of Physics, P.O. Box 64,
 FIN-00014  University of Helsinki, Finland}

%
%

\begin{abstract}

Good geometrical calibration is essential in the use of high resolution
detectors.  The individual sensors in the detector have to be calibrated with
an accuracy better than the intrinsic resolution, which typically is of the
order of $10\,\um$. We present an effective method to perform fine calibration
of sensor positions in a detector assembly consisting of a large number of
pixel and strip sensors. Up to six geometric parameters, three for location and
three for orientation, can be computed for each sensor on a basis of particle
trajectories traversing the detector system. The performance of the method is
demonstrated with both simulated tracks and tracks reconstructed from
experimental data. We also present a brief review of other alignment methods 
reported in the literature.

\end{abstract}

\maketitle

\thispagestyle{fancy}


%

\section{INTRODUCTION}

For full exploitation of high resolution position sensitive detectors, it is
crucial to determine the detector location and orientation to a precision
better than their intrinsic resolution.  It is a very demanding task to
assemble a large number of detector units in a large and complex detector
system to this high precision. Also, after assembly, the position determination
of the modules by optical  survey has its limitations because of detectors
obscuring each other. Therefore the final tuning of detector and sensor
positions is made by using reconstructed tracks.

\begin{figure}[h]
 \begin{center}
  \includegraphics[width=70mm]{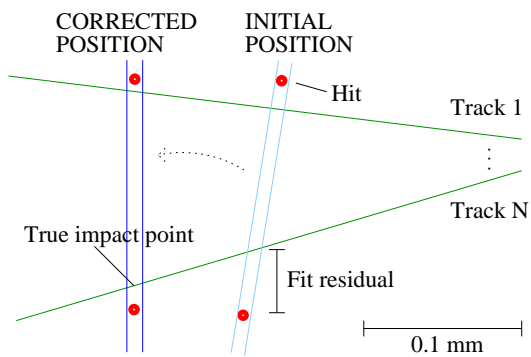}
 \caption{Schematic illustration of the method to correct the sensor 
          position: using a large number of tracks $i, i=1,\dots,N$
	  and measured hits on a detector, the sensor is
	  moved such as to minimize the residuals.} \label{fig.illustr}
 \end{center}
\end{figure}

In this paper we present an effective method by which individual sensors
in a detector setup can be aligned to a high precision with respect to each
other. The basic idea is illustrated in Figure \ref{fig.illustr}. Using a
large number of tracks, an optimum of each sensor position and orientation
is determined such that the track fit residuals are minimized.  

The outline of this paper is as follows: In Section \ref{sec.literature} we
briefly review published alignment methods. In Section \ref{sec.coordsys} we
introduce the basic notations and coordinate systems involved in our method. In
Section \ref{sec.description} we present the detailed formulation of the
method. In Sections \ref{sec.testbeam} and \ref{sec.simulation}  we demonstrate
the performance of the method applied to a test beam setup and to a simulated
pixel vertex detector, respectively. The CMS \cite{CMS-tracker} Pixel detector is
is used as a model in the simulation.

\section{BRIEF REVIEW OF ALIGNMENT METHODS} \label{sec.literature}

Most HEP experiments equipped with precise tracking detectors have to deal with
misalignment issues, and several different approaches for alignment by
tracks have been used and reported.  Most methods are iterative with
5-6 parameters solved at a time.

Several papers concerning different aspects of alignment in the DELPHI
experiment can be found in the literature. For instance, $Z^0
\rightarrow \mu^+ \mu^-$ and cosmic rays are used for the global alignment
between sub-detectors VD, OD and TPC~\cite{litDelphiGlob}.  The most
detailed DELPHI alignment paper deals with the alignment of the
Microvertex detector~\cite{litDelphiPed}.

In the ALEPH experiment, alignment was carried out wafer by wafer, and
with 20 iterations and 20000 $Z^0\rightarrow q \bar{q}$ and 4000 $Z^0
\rightarrow \mu^+ \mu^-$ events an accuracy of a few $\mu m$ can be
achieved~\cite{litAleph}.

A different, computationally challenging approach is chosen in the SLD
experiment, where the algorithm requires simultaneous solution of 576
parameters leading to a 576 by 576 matrix inversion~\cite{litSLD576}.  
In the SLD vertex detector, a recently
developed matrix singular value decomposition technique is also
used for internal alignment~\cite{litSLDsvd}.

\section{COORDINATE SYSTEMS AND TRANSFORMATIONS} \label{sec.coordsys}

Our method is applicable to detector setups which consist of planar sensors
like silicon pixel or strip detectors. For track reconstruction one conventionally
uses the local (sensor) coordinate system and the global detector system.
The local system $(u,v,w)$ is
defined  with respect to a detector module (sensor) as follows: The origin is at the
center of the sensor, the $w$-axis is normal to the sensor, the $u$-axis is along
the precise coordinate and the $v$-axis along the coarse coordinate. The global
coordinates are denoted as $(x,y,z)$.

The transformation from the global to the local system goes as:
\begin{equation}
     \bfq = \bfR(\bfr - \bfr_0)\label{transf1}
\end{equation}
where $\bfr = (x,y,z)$, $\bfq = (u,v,w)$, $\bfR$ is a rotation and 
$\bfr_0 = (x_0,y_0,z_0)$ is the position of the detector center
in global coordinates.

In the very beginning of the experiment the rotation $\bfR$ and the
position $\bfr_0$ are determined by detector assembly and survey
information. In the course of experiment this information will be
corrected by an incremental rotation  $\bfdR$ and translation  $\bfdr$
so that the new rotation and translation become:
\begin{eqnarray}
   \bfR   &\rightarrow& \bfdR\bfR \label{corr_R}\\
   \bfr_0 &\rightarrow& \bfr_0 + \bfdr.\label{corr_r}
\end{eqnarray}
The correction matrix $\bfdR$ is expressed as:
\begin{equation}
  \bfdR = \bfR_\gamma\bfR_\beta\bfR_\alpha
\end{equation}
where $\bfR_\alpha, \bfR_\beta$ and $\bfR_\gamma$ are small rotations by  $\Da,
\Db, \Dg$ around the $u$-axis, the (new) $v$-axis and the (new) $w$-axis,
respectively. The position correction $\bfdr$ transforms to the local system
as:
\begin{equation}
   \bfdq = \bfdR\bfR \bfdr\label{tr_inv}
\end{equation}
with $\bfdq = (\Delta u,\Delta v, \Delta w)$.  Using
(\ref{transf1}-\ref{tr_inv}) we find the corrected transformation from
global to local system as:
\begin{equation}
   \bfq^c = \bfdR\bfR(\bfr-\bfr_0)-\bfdq.\label{trans2}
\end{equation}
where the superscript $c$ stands for 'corrected'.   
The task of the alignment procedure by tracks is to determine the 
corrective rotation $\bfdR$ and translation $\bfdr$ or $\bfdq$ as precisely as
possible for each individual detector element.

\section{DESCRIPTION OF THE ALIGNMENT ALGORITHM} \label{sec.description}

\subsection{Basic Formulation}
Since the alignment corrections are small, the fitted trajectories can
be approximated with a straight line in a vicinity of the detector
plane. The size of this small region is determined by the alignment
uncertainty which is expected to be at most a few hundred microns so 
that the straight line approximation is perfectly valid.

The equation of a straight line in global coordinates, approximating the
trajectory in a vicinity of the detector, can be written as:
\begin{equation}
    \bfr_s(h) = \bfr_x + h\ \hbfs\label{str_unc}
\end{equation}
where $\bfr_x$ is the trajectory impact point on the detector in
question, $\hat\bfs$ is a unit vector parallel to the line and $h$ is
a parameter. Equation (\ref{str_unc}) is for {\em uncorrected}
detector positions.
 
Using Eq. (\ref{trans2}) the {\em corrected} straight line equation in the 
local system reads:
\begin{equation}
  \bfq_s(h) = \bfR_c(\bfr_x+h\ \hbfs - \bfr_0) - \bfdq\label{loc_str}
\end{equation}
where $\bfR_c=\bfdR\bfR$. A point $\bfq_s=\bfq_s(h_x)$ which lies  in the
detector plane must fulfill the condition $\bfq_s\cdot\hbfw=0$, where
$\hbfw=(0,0,1)$ is normal to the detector. From this condition we can solve the
parameter $h_x$ which gives the {\em corrected} impact or x-ing point on the
detector:
\begin{equation}
  h_x = \frac{[\bfdq-\bfR_c(\bfr_x-\bfr_0)]\cdot\hbfw}
             {\bfR_c\hbfs\cdot\hbfw}.\label{sol_t}
\end{equation}
The corrected impact point coordinates $\bfq_x^c$ in the local system are then:
\begin{equation}
  \bfq_x^c = \bfR_c(\bfr_x-\bfr_0) + 
   \frac{[\bfdq-\bfR_c(\bfr_x-\bfr_0)]\cdot\hbfw}
             {\bfR_c\hbfs\cdot\hbfw}\bfR_c\hbfs-\bfdq.\label{eq.qxc}
\end{equation}
Since the uncorrected impact point is $\bfq_x=\bfR(\bfr_x-\bfr_0)$, Eq.
(\ref{eq.qxc}) can be written as:
\begin{equation}
  \bfq_x^c = \bfdR\,\bfq_x + 
   \frac{(\bfdq-\bfdR\,\bfq_x)\cdot\hbfw}
             {\bfdR\hbft\cdot\hbfw}\bfdR\,\hbft-\bfdq.\label{eq.qxcl}
\end{equation}
where $\hbft=\bfR\hbfs$ is the uncorrected trajectory direction in the
detectors local frame of reference. Eq.~(\ref{eq.qxcl}) evaluates to:
\begin{equation}
  \bfq_x^c = \bfdR\,\bfq_x+
     (\Delta w - [\bfdR\,\bfq_x]_3)
     \frac{\bfdR\,\hbft}{[\bfdR\,\hbft]_3} -\bfdq.\label{eq.qxcll}
\end{equation}
This expression  provides us with a 'handle' by which the unknowns
$\bfdq$ and $\bfdR$ can be estimated by minimizing a respective $\chi^2$ function
using a large number of tracks.
 
\subsection{General $\chi^2$ Solution}
We denote a measured point in local coordinates as $\bfq_m=(u_m,v_m,0)$.  The
corresponding trajectory impact point is $\bfq_x^c=(u_x,v_x,0)$. For simplicity
we omit the superscripts $c$ in the coordinates $u_x$ and $v_x$. In stereo and
pixel detectors we have two measurements, $u_m$ and $v_m$, and in non-stereo
strip detectors only one, $u_m$. In the latter case the coarse coordinate $v_m$
is redundant. The residual is either a 2-vector:
\begin{equation}
    \vepsb=\left(\begin{array}{c} \veps_u \\ \veps_v\end{array}\right)
          =\left(\begin{array}{c} u_x-u_m\\v_x-v_m\end{array}\right)
	  \label{eq.epsilon}
\end{equation}
or a scalar $\veps=\veps_u=u_x-u_m$. In the following we treat the more
general 2-vector case. The scalar case is a straightforward
specification of the 2-vector formalism.

The $\chi^2$ function to be minimized for a given detector is:
\begin{equation}
  \chi^2 = \sum_j\vepsb_j^T\bV_j^{-1}\vepsb_j
\end{equation}
where the sum is taken over the tracks $j$. $\bV_j$ is the covariance matrix 
of the measurements $(u_m,v_m)$ associated with the track $j$. The alignment 
correction coefficients, i.e.~the three position parameters 
$(\Delta u, \Delta v, \Delta w)$ and the three orientation parameters 
$(\Da, \Db, \Dg)$ are found iteratively by a general $\chi^2$ 
minimization procedure. At each step of the iteration one uses the so 
far best estimate of the alignment parameters in the track fit.

Let us denote these parameters as
$\bfp =(\Delta u,\Delta v,\Delta w,\Da,\Db,\Dg)$. 
Then, according to the general $\chi^2$ solution,  the iterative correction 
to $\bfp$ has the following expression:
\begin{equation}
   \delta{\bf p} = 
    \left[\sum_j\bJ_j^T \bV_j^{-1}\bJ_j\right]^{-1}
    \left[\sum_j \bJ_j^T\bV_j^{-1}\vepsb_j\right]
\end{equation}
where $\bJ_j$ is a Jacobian matrix of $\vepsb_j(\bfp)$:
\begin{equation}
    \bJ_j = \nabla_\bfp\ \vepsb_j(\bfp). \label{nabla}
\end{equation}
An adequate starting point for the iteration is a null correction vector
$\bfp$={\bf0}.

In the general case of two measurements $(u_m,v_m)$, $\bJ_j$ is a $6\times2$
matrix. In case of scalar $\veps$, for single sided strip detectors, $\bJ_j$
is a vector of 5 elements, because $\Delta v$ is redundant and cannot be
fitted. It will also be foreseen that only a sub-set of the 6 alignment
parameters would be fitted and the others kept fixed. In this case the
dimension of the Jacobian matrix reduces accordingly.

The derivatives of the Jacobian matrix can be computed to a good precision
in the small correction angle approximation (see below). The elements of
the matrix $\bJ$ for a given track are then:
\begin{equation}
  \bJ = \left(\begin{array}{cc}
    -1       &     0        \\ 
     0       &    -1        \\
   \tanu     &  \tanv     \\ 
  v_x\tanu   &  v_x\tanv  \\
  u_x\tanu   &  u_x\tanv  \\
  v_x        & -u_x           
  \end{array}\right)
\end{equation}
The quantities $\tanu$ and $\tanv$ are defined in the next section.

\subsection{Linearized Solution with the Tilt Formalism}

We call "tilts" the angle corrections $x$ which are small enough to justify the
approximations $\cos{x}\simeq 1$  and $\sin{x}\simeq x$. In this approximation the
correction matrix $\bfdR$ reads:
\begin{equation}
   \bfdR=\left(\begin{array}{ccc} 
     1 & \Dg & \Db \\ -\Dg & 1 & \Da \\ -\Db & -\Da & 1 \end{array}\right)
     \label{eq.tiltm}
\end{equation}
Using Eq. (\ref{eq.tiltm}) we linearize Eq. (\ref{eq.qxcll}) and get the
following expressions for the corrections of the impact point  coordinates as a
function of the alignment correction parameters:
\begin{eqnarray}
  \Delta u_x &=& -\Delta u+\delta\,\tanu+\Dg\,v_x\label{eq.delux}\\
  \Delta v_x &=& -\Delta v+\delta\,\tanv-\Dg\,u_x\label{eq.delvx}
\end{eqnarray}
where $\delta = \Delta w +\Db\, u_x + \Da\, v_x$. The quantity $\angu$ is the
angle between the track and the $vw-$plane and  $\angv$ is the angle between
the track and the $uw-$plane:  $\tanu=\hat t_1/\hat t_3$, $\tanv=\hat t_2/\hat
t_3$.

With this approximation the residuals (\ref{eq.epsilon}) depend linearly on
all 6 parameters. Hence the $\chi^2$ minimization problem is linear and
can be solved by standard techniques without iteration.

\section{ALIGNMENT OF A TEST BEAM SETUP} \label{sec.testbeam}

From Eqs.~(\ref{eq.delux}) and (\ref{eq.delvx}) we can estimate the
contributions of various misalignments to the hit measurement errors.
For example the contribution of a misalignment $\Da$ around the $u$-axis 
to the $v$-coordinate is:
\begin{equation}
     \Delta v \simeq v\,\Da\,\tanv. \label{eq.dv}
\end{equation}
The error is small near normal incident angles, but grows rapidly as a function
of  $\angv$. At $\angv=45^o$ and near the edge of the sensor ($v=3$\,cm) the
error goes as $30000\,\um\,\Da$ so that for only $1\,\mrad$ error in $\Da$ the
systematic error in the $v$-coordinate is $30\,\um$.

\begin{figure}[h]
 \begin{center}
 \includegraphics[width=80mm]{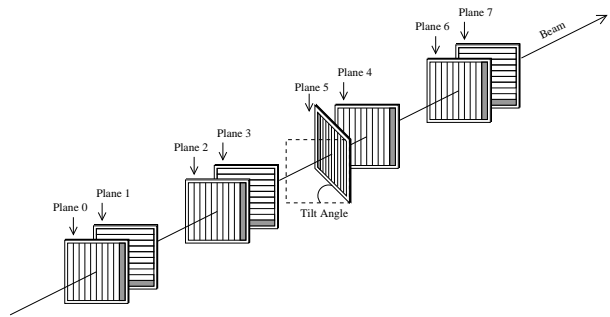}
 \end{center}
 \caption{Helsinki Si Beam Telescope in the CERN H2 beam.}\label{fig.sibt}
\end{figure}

The silicon detector team of Helsinki Institute of Physics made a precision
survey of detector resolution as a function of the angle of incidence of the
tracks \cite{calib_paper}.  The study was made in the CERN H2 particle beam
with a setup described in Figure~\ref{fig.sibt}. One of the silicon strip
detectors was fixed on a rotative support which allowed the tracks to enter
between $0$ and $30$ degrees of incident angle. The angular dispersion of the
beam was about 10\,mrad and the hits covered the full area of the test
detector. 

In order to obtain reliable results it was extremely important to calibrate the
tilt angle to a very high precision. Our algorithm was used in the alignment
calibration. In Table \ref{tab.align} we show the result of the alignment
demonstrating the  precision obtained by about 3000 beam tracks. 
\begin{table}[h]
 \begin{center}
  \caption{Alignment parameters obtained by the algorithm} \label{tab.align}
 \begin{tabular}{|c|c|c|}
  \hline
  Parameter        & At 0 degrees & At 30 degrees \\ \hline
  $\Delta u (\um)$          & 186.0$\pm$0.1   &  -264.7$\pm$0.1 \\
  $\Delta w (\um)$          & 200$\pm$20      &  -131$\pm$6 \\
  $\Delta\alpha (\mrad)$    &  5.6$\pm$0.7    &  12.9$\pm$0.9\\
  $\Delta\beta  (\mrad)$    &  5.8$\pm$0.9    &  32.59$\pm$0.04\\
  $\Delta\gamma (\mrad)$    & -14.12$\pm$0.01 & -15.86$\pm$0.01 \\
  \hline
 \end{tabular}
 \end{center}
 \end{table}

 With the precise alignment we have been able to determine the optimal
 track incident angle which minimizes the detector resolution 
 \cite{calib_paper}.



\section{MONTE CARLO SIMULATION} \label{sec.simulation}

\subsection{Simulated Detector} \label{sec.simodel}
A  Monte Carlo simulation code was written to test the alignment algorithm.
High momentum tracks were simulated and driven through a set of detector
planes. The simulated hits were fluctuated randomly to simulate measurement
errors. Gaussian multiple  scattering was added quadratically using the Highland
\cite{Highland} approximation. The algorithm involves misalignment of a
detector setup in order to simulate a realistic detector.

The experimenters' imperfect knowledge of the true position of the detector
planes is simulated by reconstructing the trajectories in the ideal (not
misaligned) detector. This means that in the transformation from local to
global coordinate system one uses the ideal positions of the detector planes.
The full algorithm in brief is as follows:
\begin{enumerate}
  \item Creation of an ideal detector setup with no misalignments
  \item Creation of a misaligned, realistic detector
  \item Generation of the particles and hits in the misaligned detector
        simulating the real detector
  \item Reconstruction of the particle trajectories in the nominal (ideal) detector
        thus using slightly wrong hit positions. This simulates the realistic
        situation in which the detector alignment is not yet performed.
\end{enumerate}

\begin{figure}[h]
 \begin{center}
  \includegraphics[width=9cm]{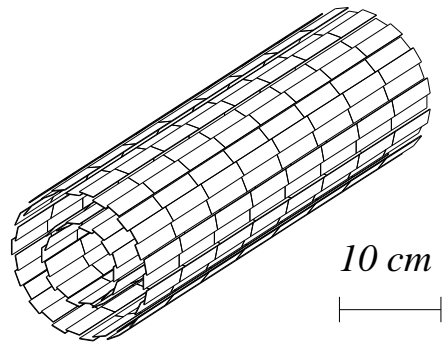}
  \caption{Illustration of the simulated vertex detector in the alignment study.}
  \label{fig.vdet}
 \end{center}
\end{figure}

For the simulated detector type we choose a vertex detector which is a
simplification of the CMS Pixel barrel detector \cite{CMS-tracker,pixel-det}
with two layers. The setup is illustrated in Figure~\ref{fig.vdet}. There are
144 sensors in layer 1 and 240 sensors in layer 2. The distance of the layer 1
from the beam line is about 4\,cm and the layer 2 about 8\,cm. 

In the simulation we used the following conditions:
\begin{enumerate}
 \item Misalignment of chosen sensors: The shifts $\Delta{u},\Delta{v},\Delta{w}$ 
       were chosen at random, each in the range $\pm100\,\um$ and the tilts 
       $\Da,\Db,\Dg$ were chosen at random each in the range $\pm{20}\,\mbox{mrad}$. 

 \item Beam and vertex constraints: The vertex positions were Gaussian
       fluctuated around the center of the beam diamond with 
       $\sigma_x=\sigma_y=20\,\um$ and $\sigma_z=7\,$cm and the tracks were
       fitted with the constraint to start from one point, i.e. from the
       primary vertex.
\end{enumerate}
In the following we consider two different cases of misaligned detectors:
\begin{enumerate}
   \item[I.] All sensors in layer 2 fixed, all sensors in layer 1 misaligned.
   \item[II.] Only one sensor in layer 2 fixed, all remaining 383 sensors
             misaligned.
\end{enumerate}

\begin{figure*}[t!]
\centering
\includegraphics[width=160mm]{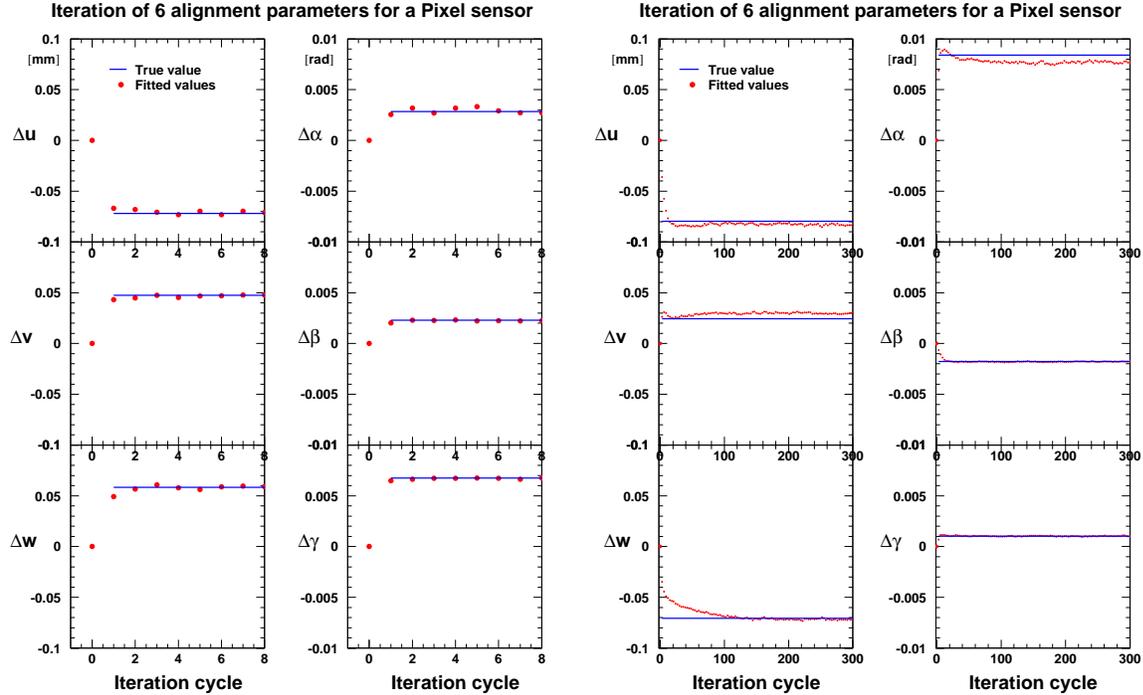}
\caption{The six plots on the left show the rate of convergence in case I for
the alignment parameters of one sensor (circular spots). The solid line shows
the true parameter value. The six plots on the right are for the case II.}
\label{fig.caseRR}
\end{figure*}

In case I the total number of fitted parameters is $6\times144=864$  and we 
used about $2\times10^5$ tracks. The case I appears to be an 'easy' one with
which the algorithm copes very well, as we see below. The second case we call
'extreme' since the alignment is based on one reference sensor which covers
only about 0.26\,\% of the detector setup area. The total number of fitted
parameters in this case was $6\times383=2298$. In the following sectios we show
perfomance results of the algorithm in these two cases.

\subsection{Convergence of the Algorithm}

The convergence rate of the alignment procedure as a function of the iteration
cycle is shown in Figure \ref{fig.caseRR}. It appears that the convergence
is fast in the 'easy' case (the 6 plots on the left) where more than 60\,\% of
the sensors provide the reference. The convergence takes place after a couple
of iterations.

In the case where only one sensor is taken as a reference (plots on the right of
the figure), the situation is different. It appears that the number of
iterations needed varies between 20 and 100 from parameter to parameter. It is
also seen that the converged parameter values are somewhat off from the true
values, but the precision is reasonable.  


\begin{figure*}[t!]
\centering
\includegraphics[width=160mm]{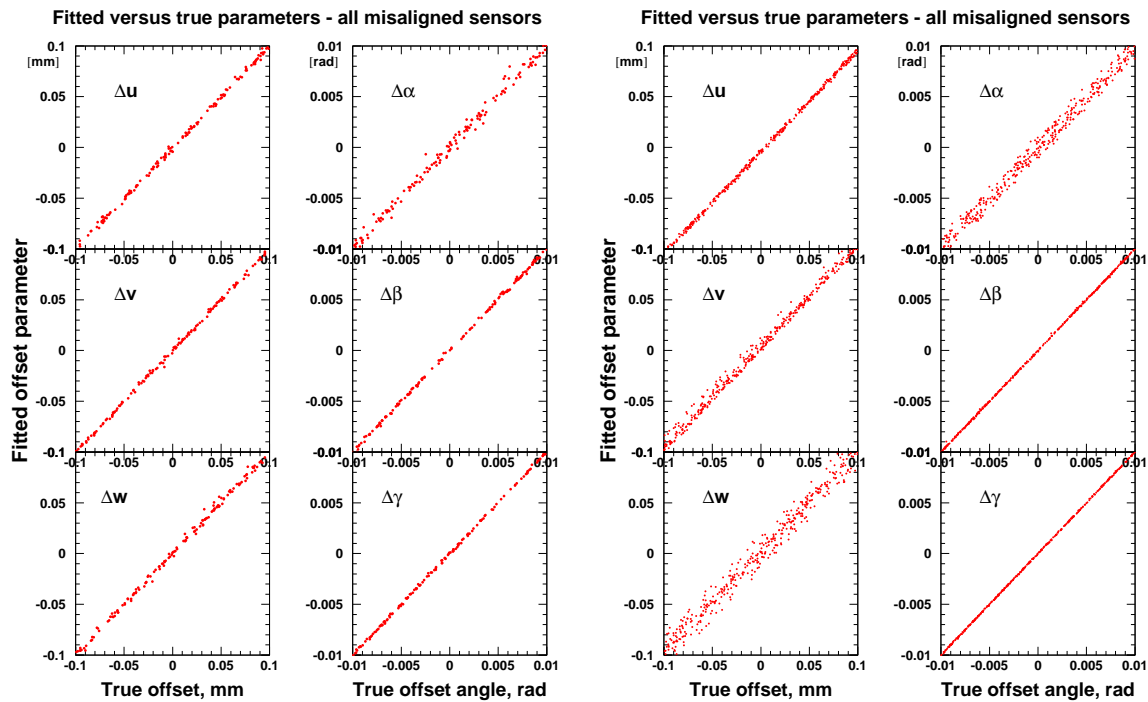}
\caption{The scatter plot of fitted versus true alignment parameters. There are
$6\times144$ entries in the plots on the left (case I) and $6\times383$ 
entries in the plots on the right (case II).}
 \label{fig.caseCC}
\end{figure*}

\subsection{Comparison of Fitted and True Parameters}
The precision of the fitted parameters in comparison with the true values is
shown in Figure \ref{fig.caseCC} on the left for the case I. The correlations
are very strong. The typical deviation of the fitted parameters from the true
value is less than $1\,\um$ for the offsets and a fraction of a milliradian
for the tilts. The precision appears to be better than actually needed in 
this case, indicating that a smaller statistics would give a satisfactory
result.

In case II (the plots on the right of the figure) a good correlation is
observed, but the precision is somewhat more modest. For example the
error in $\Delta w$ (the shift normal to the sensor plane) is still in most
cases below $10\,\um$.  

 
\section{CONCLUSIONS} \label{sec.conclusions}

We have developed a sensor alignment algorithm which is mathematically and 
computationally simple. It is based on repeated track fitting and residuals
optimization by $\chi^2$ minimization. The computation is simple, because the
solution involves matrices whose dimension is at most $6\times6$. The method is
capable of solving simultaneously all six alignment parameters per sensor for a
detector setup with a large number of sensors. 

We have successfully applied the method in a precision survey of silicon strip
detector resolution as a function of the tracks incident angle. Furthermore, we
have demonstrated the performance of the algorithm in case of a simulated
two-layer pixel barrel vertex detector. The method performs very well in the
case where the outer layer is taken as a reference and all inner sensors are to
be aligned. The algorithm performs reasonably well also in the extreme case
where only one sensor, representing some 0.26\,\% of the total area, is
taken as a reference for the alignment. \\

\begin{acknowledgments}
The authors wish to thank K.~Gabathuler, R.~Horisberger and D.~Kotlinski for inspiring
discussions. 

Work supported by Ella and Georg Ehrnrooth foundation, Magnus Ehrnrooth
foundation, Arvid and Greta Olins  fund at Svenska kulturfonden
and Graduate School for Particle and Nuclear Physics, Finland.
\end{acknowledgments}

\end{document}